\documentstyle [cite,rotating,epsf,twoside,11pt,here]{article}

\def\pt  {$p_{T}$}
\def\xf  {$x_{F}$}

\def\eplus {$e^{+}$}
\def\emin {$e^{-}$}

\def\piplus {$\pi^{+}$}
\def\pimin {$\pi^{-}$}
\def\pizero {$\pi^{0}$}

\def\kplus {K$^{+}$}
\def\kmin {K$^{-}$}
\def\kzero {K$^{0}$}

\def\sigmaplus {$\Sigma^{+}$}
\def\sigmamin {$\Sigma^{-}$}
\def\sigmazero {$\Sigma^{0}$}

\def\lambdazero {$\Lambda^{0}$}
\def\antilambda {$\overline{\Lambda}^{0}$}

\def\xizero {$\Xi^{0}$}
\def\ximin {$\Xi^{-}$}

\def\omegamin {$\Omega^{-}$}

\def\xicharmplus {$\Xi_{c}^{+}$}
\def\xicharmzero {$\Xi_{c}^{0}$}

\def\omegacharmzero {$\Omega_{c}^{0}$}

\def\lambdacharmplus {$\Lambda_{c}^{+}$}

\def\kevc1{\ifmmode\mathrm{\ keV/{\mit c}}
          \else$\mathrm{\ keV/{\mit c}}$\fi}
\def\Mevc1{\ifmmode\mathrm{\ MeV/{\mit c}}
          \else$\mathrm{\ MeV/{\mit c}}$\fi}
\def\gevc1{\ifmmode\mathrm{ GeV/{\mit c}}
          \else$\mathrm{ GeV/{\mit c}}$\fi}
\def\kevc2{\ifmmode\mathrm{\ keV/{\mit c}^2}
          \else$\mathrm{\ keV/{\mit c}^2}$\fi}
\def\Mevc2{\ifmmode\mathrm{\ MeV/{\mit c}^2}
          \else$\mathrm{\ MeV/{\mit c}^2}$\fi}
\def\Gevc2{\ifmmode\mathrm{\ GeV/{\mit c}^2}
          \else$\mathrm{\ GeV/{\mit c}^2}$\fi}
\def\Gev2c2{\ifmmode\mathrm{\ GeV^2/{\mit c}^2}
          \else$\mathrm{\ GeV^2/{\mit c}^2}$\fi}

\def\Pp{\ifmmode{\mathrm p}
         \else${\mathrm p}$\fi}
\def\Pn{\ifmmode{\mathrm n}
         \else${\mathrm n}$\fi}

\begin{document}
\begin{flushright}
CERN PPE 97-23 \\
MPI-K H V9 1997 \\*[8mm]
19 February 1997
\end{flushright}
\vspace{0.8cm}
\Large
           
\centerline{$\Xi ^{-}$ Production by $\Sigma ^{-}$,$\pi^{-}$ and Neutrons in the }
\centerline{Hyperon Beam Experiment at CERN}
\large
\vspace{1.8cm}
\centerline{The WA89 Collaboration}
\vspace{0.4cm}
\normalsize

\sloppy
  M.I.~Adamovich$^8$, 
  Yu.A.~Alexandrov$^8$,
  D.~Barberis$^3$, 
  M.~Beck$^5$, 
  C.~B\'erat$^4$,
  W.~Beusch$^2$, 
  M.~Boss$^6$,
  S.~Brons$^{5,f}$, 
  W.~Br\"uckner$^5$, 
  M.~Bu\'enerd$^4$, 
  C.~Busch$^6$,
  C.~B\"uscher$^5$,
  F.~Charignon$^4$, 
  J.~Chauvin$^4$, 
  E.A.~Chudakov$^{8,h}$, 
  U.~Dersch$^5$, 
  F.~Dropmann$^5$, 
  J.~Engelfried$^{6,a}$, 
  F.~Faller$^{6,b}$, 
  A.~Fournier$^4$,
  S.G.~Gerassimov$^{5,7}$, 
  M.~Godbersen$^{5}$, 
  P.~Grafstr\"om$^2$,
  Th.~Haller$^5$,
  M.~Heidrich$^5$, 
  E.~Hubbard$^{5}$, 
  R.B.~Hurst$^3$, 
  K.~K\"onigsmann$^{5,c}$,
  I.~Konorov$^5$, 
  N.~Keller$^6$,
  K.~Martens$^{6,g}$,
  Ph.~Martin$^4$, 
  S.~Masciocchi$^{5}$, 
  R.~Michaels$^{5,d}$, 
  U.~M\"uller$^7$,
  H.~Neeb$^{5}$,
  D.~Newbold$^1$, 
  C.~Newsom$^{e}$, 
  S.~Paul$^5$,
  J.~Pochodzalla$^5$,
  I.~Potashnikova$^5$, 
  B.~Povh$^5$,
  Z.~Ren$^5$, 
  M.~Rey-Campagnolle$^4$,
  G.~Rosner$^7$,
  L.~Rossi$^3$, 
  H.~Rudolph$^{7}$,
  C.~Scheel$^i$, 
  L.~Schmitt$^7$, 
  H.-W.~Siebert$^6$, 
  A.~Simon$^{6,c}$,
  V.~Smith$^1$,
  O.~Thilmann$^6$, 
  A.~Trombini$^{5}$, 
  E.~Vesin$^4$, 
  B.~Volkemer$^{7}$,
  K.~Vorwalter$^5$, 
  Th.~Walcher$^7$, 
  G.~W\"alder$^{6}$, 
  R.~Werding$^{5}$,
  E.~Wittmann$^5$,
  M.V.~Zavertyaev$^{5,8}$ 

\vspace{0.2cm}
\begin{flushleft}
\noindent
    $^1${\sl University of Bristol, Bristol, United Kingdom} \\
    $^2${\sl CERN; CH-1211 Gen\`eve 23, Switzerland.}\\
    $^3${\sl Genoa Univ./INFN; 
            Dipt. di Fisica, Via Dodecaneso 33, 
             I-16146 Genova, Italy.}\\
    $^4${\sl Grenoble ISN; 
          53 Avenue des Martyrs; CEDEX, 
          F-38026 Grenoble, France.}\\
    $^5${\sl Heidelberg Max-Planck-Inst. f\"ur Kernphysik$^\#$; 
          Postfach 103980, 
          D-69029 Heidelberg, Germany.}\\
    $^6${\sl Heidelberg Univ., Physikal. Inst.$^\#$; 
          Philosophenweg 12, 
         D-69120 Heidelberg, Germany.}\\
    $^7${\sl Mainz Univ., Inst. f\"ur Kernphysik$^\#$; 
          Johann Joachim Becher Weg 45,
          D-55099 Mainz, Germany.}\\
    $^8${\sl Moscow Lebedev Physics Inst.; 
          Leninsky Prospect 53, 
          RU-117924, Moscow, Russia.}\\
\end{flushleft}
\vspace{0.2cm}

\normalsize
\centerline{\underline{\large{Abstract}}}
\vspace{0.5cm}

Inclusive cross sections for \ximin\ hyperon production in 
high-energy \sigmamin, \pimin\ and neutron induced interactions
were measured by the experiment WA89 at CERN.
Secondary  \sigmamin\ and \pimin\ beams with average momenta of 345~\gevc1
and a neutron beam  of average momentum 65~\gevc1 were produced 
by primary protons of 450~\gevc1\ from the CERN SPS.
Both single and double differential cross sections are presented
as a function of the transverse momentum and the Feynman variable $x_F$.
A strong leading effect for \ximin\ produced by \sigmamin\ is
observed. The influence of the target mass on the \ximin\ cross section 
is explored by comparing reactions on copper and carbon nuclei.
\vspace{0.5cm}

\centerline{(Submitted to Zeitschrift f\"ur Physik C)}
\vspace{2.0cm}

\hrule width 6truein
\vspace{0.2cm}
\small
\noindent
{$\#$) supported by the Bundesministerium f\"ur Bildung,
  Wissenschaft,Forschung und Technologie, 
  Germany, under contract numbers 05~5HD15I, 06~HD524I and 06~MZ5265 }

\newpage

\setlength{\oddsidemargin}   {0.0in}
\setlength{\evensidemargin}  {0.0in}
\setlength{\textwidth}       {6.5in}
\setlength{\textheight}      {9.0in}
\setlength{\topmargin}      {-0.3in}
\setlength{\headheight}      {0.3in}
\setlength{\headsep}         {0.3in}
\setlength{\footskip}        {0.6in}
\setlength{\footheight}      {0.3in}
\vspace{4in}
\hrule width 2truein
\vspace{0.15cm}
\small
\noindent
    $^a${\sl Now at FNAL, PO Box 500 Batavia, 
          IL 60510, USA.} \\
    $^b${\sl Now at Fraunhofer Inst. f\"ur Solar Energiesysteme,
               D-79100 Freiburg, Germany.}\\
    $^c${\sl Now at Fakult\"at f\"ur Physik, Univ. Freiburg, Germany}\\
    $^d${\sl Now at Thomas Jefferson Lab , 12000 Jefferson Ave., 
          Newport News, VA 23606, USA.} \\
    $^e${\sl University of Iowa, Iowa City, IA 52242, USA.}\\
    $^f${\sl Now at Triumf, Vancouver, B.C., Canada V6T 2A3 }\\
    $^g${\sl Now at Kamioka Observatory, 
             Institute for Cosmic Ray Research, University of
             Tokyo,Japan} \\
    $^h${\sl On leave from Moscow State University, Moscow, Russia}\\
    $^i${\sl Nikhef, 1009 DB Amsterdam, The Netherlands }

\normalsize

\clearpage

\pagestyle{plain}



\def\eplus {$e^{+}$}
\def\emin {$e^{-}$}

\def\piplus {$\pi^{+}$}
\def\pimin {$\pi^{-}$}
\def\pizero {$\pi^{0}$}

\def\kplus {K$^{+}$}
\def\kmin {K$^{-}$}
\def\kzero {$K^{0}$}

\def\sigmaplus {$\Sigma^{+}$}
\def\sigmamin {$\Sigma^{-}$}
\def\antisigmaplus {$\overline{\Sigma}^{-}$}
\def\sigmazero {$\Sigma^{0}$}

\def\lambdazero {$\Lambda^{0}$}
\def\antilambda {$\overline{\Lambda}^{0}$}

\def\xizero {$\Xi^{0}$}
\def\ximin {$\Xi^{-}$}
\def\antiximin {$\overline{\Xi}^{+}$}

\def\omegamin {$\Omega^{-}$}

\def\xicharmplus {$\Xi_{c}^{+}$}
\def\xicharmzero {$\Xi_{c}^{0}$}

\def\omegacharmzero {$\Omega_{c}^{0}$}

\def\lambdacharmplus {$\Lambda_{c}^{+}$}

\def\pt  {$p_{T}$}
\def\xf  {$x_{F}$}


\setcounter{page}{1}
\section{Introduction}
\label{sec_1}
The production of strange or heavy quarks and their subsequent hadronization
in hadron-hadron collisions constitute an important benchmark
test for QCD-inspired phenomenological models 
\cite{TAK83,TAS87,TAS88,NAZ89,NAZ92} describing soft phenomena and 
the applicability of perturbative QCD to hard processes.
Using, for example, projectiles with different flavour contents,
the role of valence quarks can be explored \cite{TAK83,TAS88,DUR96}.
In addition, hyperons represent an essential ingredient for our understanding
of hadronic systems at high energy density.
In reactions with complex nuclei the production of strange 
\cite{RAF82,KOC86} and charmed \cite{MAT86,GEI93,VOG94,LEV95,ZI95,SHU96,HWA96}
particles is sensitive to the surrounding nuclear medium.
In turn, the produced particles may allow to probe interactions with 
and properties of the partonic or hadronic medium itself.
Evidence for unusual production mechanisms in a non-hadronic 
environment will always be based on a comparison to
hadron-hadron and hadron-nucleus data.   
Finally, on a macroscopic scale, hyperons are predicted to play a
relevant role during the developement of neutron stars 
\cite{GLE82,GLE87,SCH96}.

The production of hyperons has been studied mostly
in proton and neutron induced reactions
\cite{BAD72,HUN75,SKU78,BOU79,ERH79,BAU81,BOU80,CAR85,BER86,GOT96,ALE86}
or with heavy ion beams \cite{ABA91,AND94,BAR95}.
Only few data exist for pions interacting with nuclear targets
\cite{FEN84,MIK86}. Despite the potential interest, measurements
with strange quarks in the incoming beam are also very scarce.
$\Xi^-$ production in $\Xi^-$ + Be and $\Xi^-$ + N interactions was measured
in earlier experiments using the charged hyperon beam at CERN
\cite{BIA81,BIA87}. Since the various experiments were performed in different 
kinematic  regions, large systematic uncertainties make a direct
comparison of these measurements rather difficult.

The latest hyperon beam experiment at CERN -- named WA89 -- 
offers the opportunity to measure the
$\Xi^-$ production under homogeneous conditions with different types
of beam on several targets. In this paper we present the first
measurement of $\Xi^-$ production
in $\Sigma^-$ interactions, thus filling the gap between non-strange
beams and the $\Xi^-$ induced reactions with two strange quarks in the
incoming projectile. In addition, we have studied the
$\pi^- A \rightarrow \Xi^- X$ and
$n A \rightarrow \Xi^- X$ reactions which enable to connect the
new data obtained in the present study with the results of previous 
measurements.

The paper is organized as follows. 
Experimental details are presented in the following section 
\ref{sec_2}. In section \ref{sec_3} the results are presented.
A comparison with existing data and a discussion follows in 
section~\ref{sec_4}.
The final conclusions and a summary are then given in section
\ref{sec_5}.
%
\section{Experimental Setup and Event Selection}
\label{sec_2}
%
The experiment WA89 was performed using the charged hyperon beam 
of the CERN SPS. 
Its main purpose is to study the production, spectroscopy and decays 
of charmed baryons and to search for exotic states. 
The production studies of \ximin\ were performed in parallel 
with the main program.
%
\subsection{Beamline and apparatus}
\label{sec_2_1}
Hyperons were produced by 450~\gevc1 protons impinging on a 40~cm long
beryllium target with a diameter of 0.2~cm.
A magnetic channel consisting of 3 magnets 
with an integrated field of 8.4~Tm selected negative particles 
with a mean momentum of 345~\gevc1, a momentum spread of $\sigma (p)/p=9\%$,
and an angle to the proton beam smaller than 0.5~mrad.
After a distance of 16~m the produced hyperons hit the
experimental target which consisted of one copper and three carbon
(diamond) blocks
arranged in a row along the beam. Each copper and carbon block
had a thickness corresponding to an interaction length 
of 2.6~\% and 0.83~\%, respctively. At the target the beam 
was homogeneously distributed
over a rectangular area with a width of 3~cm and 
a height of 1.7~cm. Its dispersion was 0.6~mrad in the horizontal plane and
1.0~mrad in the vertical plane.
An average beam spill of 2.1~s contained about $1.8\cdot 10^{5}$
\sigmamin\ hyperons and about $4.5\cdot 10^{5}$ \pimin\ at the 
experimental target for an incoming
intensity of $4.0\cdot 10^{10}$ protons per spill. A transition radiation
detector (TRD) \cite{BRU96} was used to discriminate online between 
\pimin\ and hyperons. In special runs \pimin\ 
interactions were recorded for normalisation purposes.

\begin{figure}[H]
  \begin{rotate}{90}
\mbox{\epsfxsize=13cm\epsffile{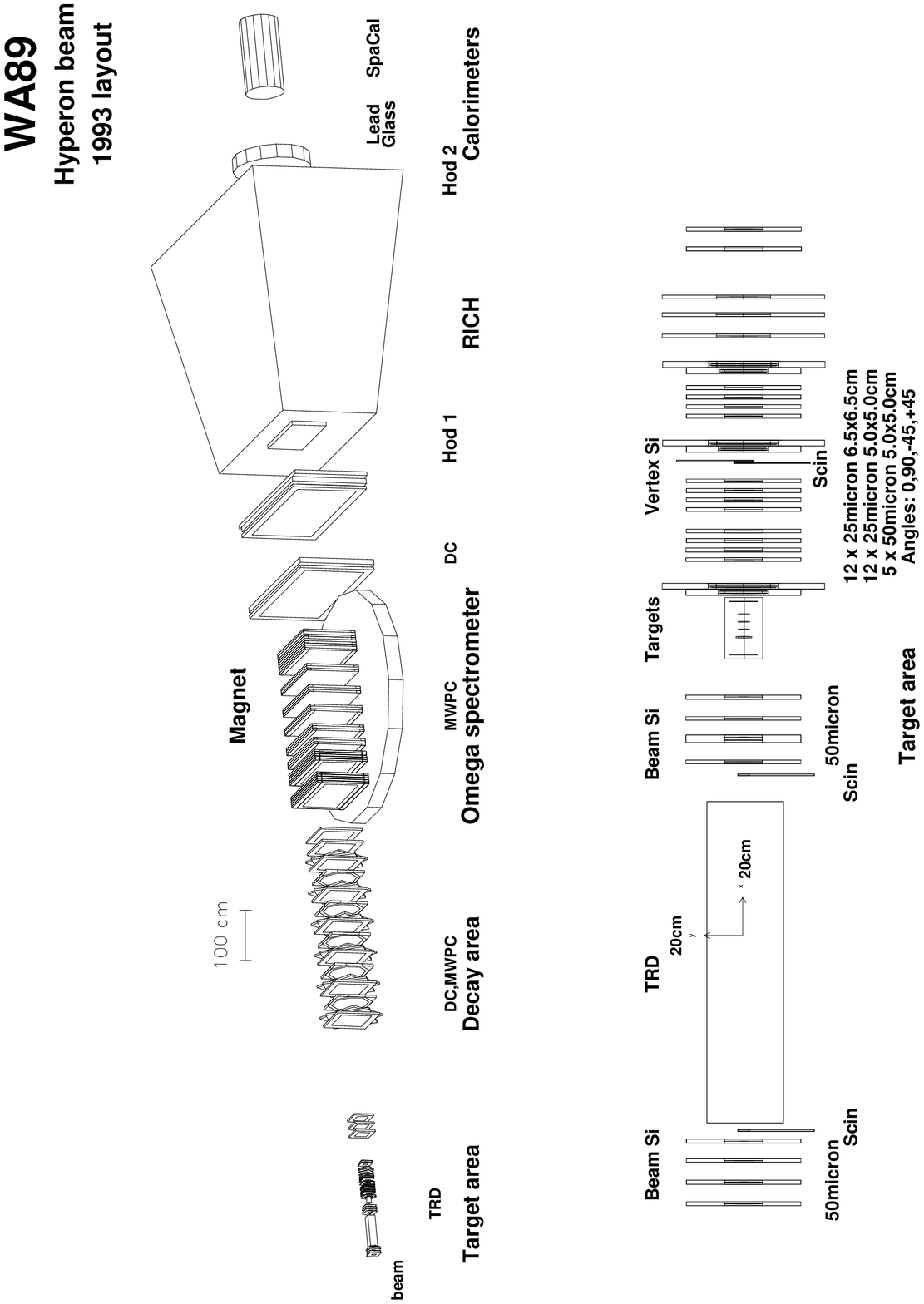}}
  \end{rotate}

  \vspace{12.7cm}
\caption{Setup of the WA89 experiment in the 1993 run.
        The lower part shows an expanded view of the target area.}
\label{fig:setup}
\end{figure}

Fig. \ref{fig:setup} shows a sketch of the experimental setup 
used in the 1993 run of WA89 on which the data of this paper are based.
The beam and the secondary particles were detected
by 29 silicon micro-strip planes with 25 and 50$\mu $m pitch.
Positioning the target about 14~m
upstream of the centre of the $\Omega $-spectrometer provided a 10~m long
decay area for short living strange particles. The products
of these decays along with the particles coming directly from the target
were detected by 40 planes of drift chambers with a spatial resolution
of about 300~$\mu $m. Special MWPC chambers (20 planes with 1mm wire spacing)  
were used in the central region of high particle fluxes.
In order to improve the track bridging between the target region and the
decay region three sets of 4 MWPCs each with a pitch of 1 mm
were installed about 2 m behind the target.

The particle momenta were measured by the
$\Omega$-spectrometer \cite{BEU77} consisting of a super-con\-duc\-ting 
magnet with a field integral of 7.5~Tm and a tracking detector
consisting of 45 MWPC planes inside the field area and 12 drift chamber
planes at the exit of the magnet. The momentum resolution was
$\sigma (p)/p^{2}\approx 10^{-4}~(\gevc1)^{-1}$.

Charged particles were identified
using a ring imaging Cherenkov (RICH) detector \cite{BEU92}. It had a 
threshold of $\gamma =42$  and provided $\pi/p$ separation
up to about 150~\gevc1. Downstream of the RICH a lead glass electromagnetic  
calorimeter was positioned for photon and electron detection \cite{BRU92}.
This calorimeter was followed by a hadron calorimeter\cite{SCH94}.

The trigger selected about 25$\% $ of all
interactions using multiplicities measured in scintillators in the
target region and in hodoscopes and proportional 
chambers downstream of the magnet. 
Correlations of hits in these detectors were used to select 
particles with high momenta thus reducing the background from low-momentum
pions (see section \ref{sec_2_3_1}). More than 2 particles
at the exit of the magnet were required by the trigger.
The results shown in the present paper are based on 
the analysis of about 100 million events recorded in 1993.
%
\subsection{Event reconstruction}
\label{sec_2_2}
The event reconstruction proceeded in two steps. In the first 
part of the analysis events with a \ximin\ candidate were
selected. Secondly, interactions in the target were
identified, with different constraints applied for the different 
beam particles.
%
\subsubsection{\ximin\ identification}
\label{sec_2_2_1}
\ximin\ were reconstructed in the decay chain 
$\Xi^- \rightarrow \Lambda^0 \pi^- \rightarrow p \pi^- \pi^- $.
Only $\Lambda^0$ -decays upstream of the magnetic field of 
the $\Omega $-spectrometer were considered in this analysis.
Reconstructed segments in the chambers of the decay area 
and in the $\Omega $-spectrometer were required for the tracks 
of the daughter particles 

Candidates for \ximin\ -decays were then selected by the following 
procedure. For the definition of the $\Lambda^0 \rightarrow p \pi ^-$ decays, 
all combinations of positive and negative tracks were considered.
The distance of the two tracks at the decay point was 
not allowed to exceed 0.5 cm.
To look for $\Xi^- \rightarrow  \Lambda ^0 \pi ^-$ decays, only 
\lambdazero\ candidates within a mass window of\ $3\sigma $ around the
reference mass were accepted. Here $\sigma$ denotes the uncertainty 
of the mass determination based on the track properties of the 
individual events. (Typically, $\sigma$ is about 3.7 \Mevc2). 
In addition, the corresponding \ximin\ trajectory 
had to be measured in the vertex detector.
The momentum spectrum of the  \ximin\ and \lambdazero\ candidates starts at
about 12~\gevc1, the cutoff being due to the spectrometer acceptance.
%
\subsubsection{Selection of interactions}
\label{sec_2_2_2}
\sigmamin\ , \pimin\ and neutron interactions were identified according to 
the following conditions. The interaction vertex contains at least two
outgoing charged tracks one of which is the \ximin\ track. 
Furthermore, the reconstructed vertex position had to be within 
a target block where in each coordinate an additional margin of 
3$\sigma$ was allowed. The $\sigma$ denotes the uncertainty of the vertex
position calculated for each individual event from the track parameters.

For \sigmamin\ and \pimin\ interactions
the transverse distance between the \sigmamin\ or \pimin\ beam track  and 
the reconstructed interaction vertex position was required to be less 
than $6 \sigma$
($\sigma \approx 25 \mu m$). Events were rejected if 
the beam track was connected to an outgoing track.

To identify interactions of neutrons generated by \sigmamin\ decays in the
beam channel the following criteria were imposed.
Since \pimin\ from \sigmamin\ decays are below the TRD threshold,
it was required that no high momentum \pimin\ was detected in the TRD.
In addition, the \pimin\ track from the \sigmamin\ decay had to pass
the reconstructed interaction point with a distance of at least $6 \sigma$.
Finally, it was demanded that the \pimin\ track was connected 
to a track in the spectrometer and that it had a momentum smaller 
than 140 \gevc1 corresponding to the $\Sigma^- \rightarrow n \pi^-$ 
decay kinematics at \mbox{$<p> \approx $ 345 \gevc1.}

\begin{figure}[H]
\begin{center}
\mbox{\epsfxsize=13cm\epsffile{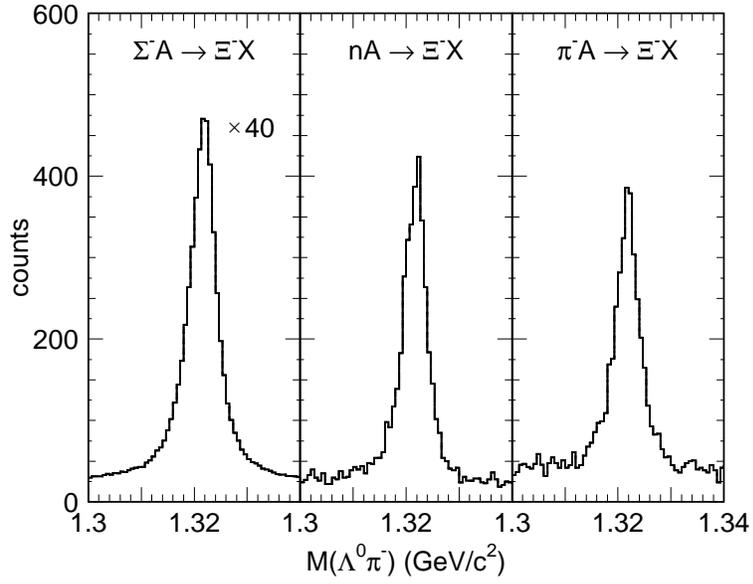}}
\caption{Invariant mass distributions of $\Lambda^0-\pi^-$ pairs
        produced in \sigmamin\ (left), neutron (centre), 
        and \pimin\ (right) interactions.
        The data for both targets have been added.
        }
\label{fig:ksy_mass}
\end{center}
\end{figure}

The resulting \ximin\ mass spectra for the different incident beam
particles are shown in Fig. \ref{fig:ksy_mass}. We observed about
178,000 \ximin\ from  \sigmamin\ interactions,  3000 \ximin\ from \pimin\ 
interactions and 3500 from neutron interactions. The mass resolution
lies in the range from  2.5 \Mevc2 to  2.8 \Mevc2
and is dominated by the error on the direction of the decay products.
%
\subsection{Beam properties}
\label{sec_2_3}
The negative beam particles ( \sigmamin\ , \pimin\ ) had an average momentum 
of 345~\gevc1 and a momentum spread of $\sigma (p)/p=9\%$. Since the beam 
momentum in each individual event was not measured, we used an average momentum 
of 345~\gevc1 for the further analysis.

Despite the beam particle identification in the TRD, the
\sigmamin or \pimin\ data taken in the present experiment 
were contaminated by misidentified \pimin,
\sigmamin, \ximin,  \kmin\ and neutrons.
For all our measurements the number of \ximin\ in the beam 
is most crucial due to the large \ximin\ production cross section in
the  reaction  $\Xi^- + A \rightarrow \Xi^- + X $ \cite{BIA81}.
Therefore the  beam composition and its influence on the cross
section measurement was 
studied in a detailed analysis for each data set individually.
\subsubsection{The \sigmamin beam}
\label{sec_2_3_1}
\begin{figure}[H]
\begin{center}
\mbox{\epsfxsize=15cm\epsffile{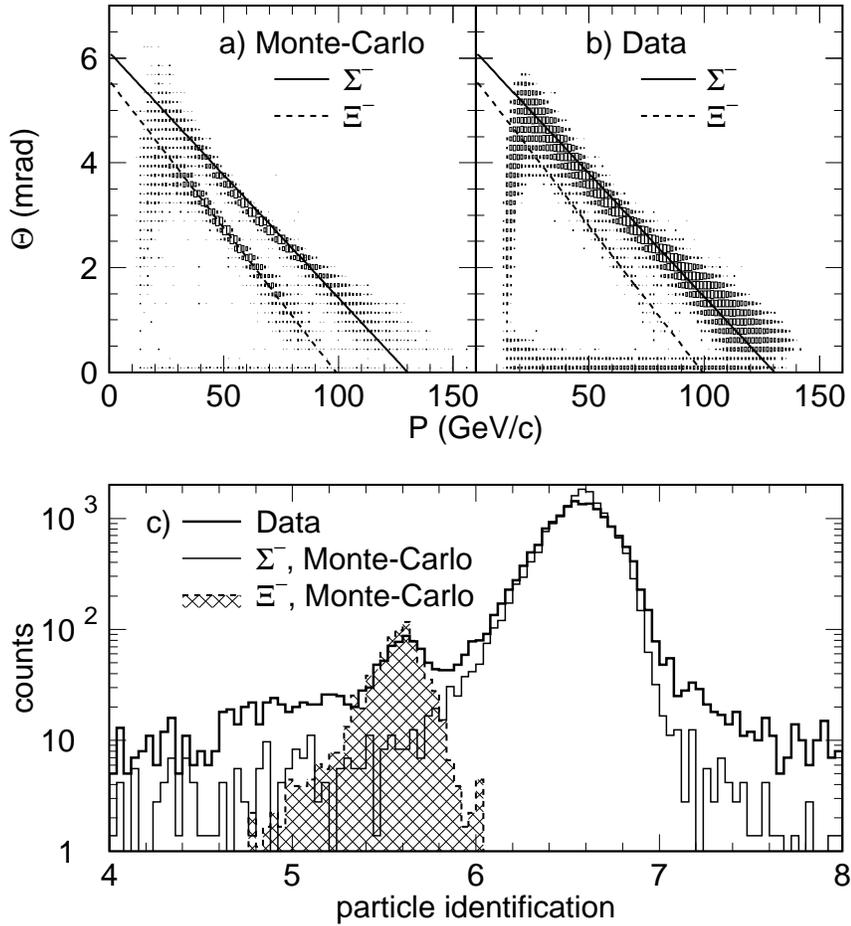}}
\caption{Correlation plots between momentum and decay angle between 
incoming \sigmamin\ and \ximin\ and outgoing \pimin\ for a) Monte
Carlo and b) Data. The projections of these distributions along the two indicated 
lines is shown in part c).
}
\label{fig:ksy_kink}
\end{center}
\end{figure}

To measure the \ximin\ contamination of the \sigmamin\ beam, a sample
of events with beam particles passing through the target without
interaction was analyzed. In this data sample we identified 
\sigmamin\ and \ximin\ decays by the
observation of the decay  kinks  between the incoming 
beam particle and a negative particle detected in the spectrometer 
from the decays $ \Sigma^- \rightarrow n \pi^- $ and
$ \Xi^- \rightarrow \Lambda \pi^- $, respectively.
The decay point was accepted if it was located between the last micro 
strip plane and the first set of drift chambers of the decay area. 
The decay lengths of \sigmamin\ and \ximin\ at the same momentum differ by
only $1\%$  and therefore  no acceptance or efficiency corrections 
are needed to evaluate the \mbox{\sigmamin\ / \ximin\ } ratio. 
In part a) of Fig.~\ref{fig:ksy_kink} the expected correlation between
the momentum and the kink angle is shown for 
\sigmamin\ and \ximin\ decays. In order to determine the 
separation between the different beam particles, an equal number of 
 \sigmamin\ and \ximin\ particles was generated 
in this Monte Carlo simulation.
The experimental correlation between the 
momentum and the kink angle is displayed in Fig.~\ref{fig:ksy_kink}b.
A second band originating from \ximin\ decays can be discerned there.
For a quantitative analysis we projected this distribution  
along the angle-momentum correlation marked by the dashed and 
solid lines. This projection is shown by the solid 
histogram in part c) of Fig.~\ref{fig:ksy_kink}. 
Besides the dominating peak from \sigmamin\ decays, a well
separated maximum corresponding to \ximin\ decays can be identified.
The relative strength of the two peaks can be estimated with the  
Monte Carlo simulations by assuming a 
\ximin\ contribution of (1.26 $\pm$ 0.07)\% to the \sigmamin\ beam 
(see cross-hatched histogram in Fig.~\ref{fig:ksy_kink}.c).
Based on the $\Xi^- + Be \rightarrow \Xi^- + X $ cross section 
from Ref.~\cite{BIA87} the differential cross sections
were corrected for this contamination.

The actual \pimin\ to \sigmamin\ ratio in the beam is
about 2.3. Although beam \sigmamin\ were identified online by the TRD, 
the data sample taken during the experiment was still contaminated 
by misidentified \pimin\ with large momenta. 
The amount of this contamination was determined offline with 
the help of the pulse height information from each chamber in the TRD. 
A value of $(12.3 \pm 0.5 )\% $ of remaining high momentum \pimin\ 
in the \sigmamin\ data sample was obtained. Using the 
$\pi^- + A \rightarrow \Xi^- + X $ 
cross section measured in the present experiment (see below),
we corrected the observed 
$\Sigma^- + A \rightarrow \Xi^- + X $ yield in an iterative way for 
these \pimin\ interactions.

The $K^-$ to \pimin\ ratio in 300 \gevc1 p-Be collisions was measured 
to be $(1.15 \pm  0.02)\% $ \cite{ATH80}.
Since high momentum $K^-$ mesons cannot be separated from 
the \sigmamin\ beam via the TRD information, 
$\rm 2.1\%$ of the total beam in our data sample consists of $K^-$. 
The production cross section for $K^-p \rightarrow \Xi^- X $ was
measured only at low beam momenta between 4.2 and  16 \gevc1 
\cite{BAU81}. Within this momentum range the cross 
section slowly decreases from $157 \pm 8 \mu b $ to $135 \pm 15 \mu b $ .
Taking  $ 157 \mu b $ as an upper limit for the cross section of
\ximin\ produced by $K^-$ at 345 \gevc1 we estimated that at most 0.4\%
of the  observed \ximin\ yield can be attributed to $K^-$ 
content of the beam. Since the actual value for the $K^-$ cross 
section is unknown we include this contamination in the systematic error.

Finally, the \sigmamin\ data sample 
was contaminated by low momentum pions and neutrons stemming from 
\sigmamin\ decays between the exit of the beam channel and the target.
Monte Carlo simulations predicted that 28\%  
of the \sigmamin\ at the exit of the beam channel decayed with 
the daughter \pimin\ hitting the beam scintillators 
and the target. Thus, slow secondary pions should acount for
about 24\% of the total flux {\it at the target}. 
According to these simulations a strong interdependence between 
the incident angle and the position at the target is expected for the
primary beam particles while 
no such correlation exists for slow, secondary pions.
Indeed a plot for the experimental data of the beam angle versus 
the target position shows  a narrow band which is superimposed 
on a homogeneous background. 
We find that 23\% of the total incoming beam
flux is contained in this background. Motivated by the good agreement of 
this number with the expected 24\% of secondary pions,
we attribute the particles outside of the narrow position-angle 
correlation band to decay pions. 
We  therefore corrected the total incident beam flux for a
23\% contribution from low momentum pions.
By cutting on the position-angle correlation  
the contamination to the observed \ximin\ yield from these 
slow pions could be suppressed.

\subsubsection{The \pimin\ beam}
\label{sec_2_3_2}

To obtain a sample of interactions of beam pions the online TRD decision was
inverted. This data sample contains misidentified 
\sigmamin, \ximin, and $K^-$-mesons. 
With the help of the  TRD offline  analysis we determined 
the total remaining contamination in the pion beam to be $(1.5  \pm 1)\%$.
In order to correct the pion data for this background, we assumed
that the relative distribution between 
\sigmamin\ , \ximin and \kmin is identical to that of the original beam 
of about 78~:~1~:~3.
Using the 
$\Sigma^- + A \rightarrow \Xi^- + X $ cross section from the present study
and the 
$\Xi^- + Be \rightarrow \Xi^- + X $ data at 116 \gevc1 \cite{BIA87},
we obtained corrections of about 10\% and
1.5\% due to the \sigmamin\ and \ximin contributions, respectively.
Because of the additional
TRD suppression, the contribution to the \pimin sample 
from the $K^-$ component is negligible.

\subsubsection{The neutron beam}
\label{sec_2_3_3}

Neutrons originating from \sigmamin\ decays upstream of the 
target were used to measure the \ximin\ production by neutrons. 
The momenta of these neutrons were defined as the difference between 
the average \sigmamin\ momentum and the momentum of the associated
\pimin\ measured in the spectrometer. 
The neutron spectrum has an average momentum of 260~\gevc1\ and a width
of $\sigma (p)/p=15\%$.

Despite the tagging of the neutron interactions by the associated
pions, this data sample is also not free of contaminations.
$\Lambda^0$ stemming from \ximin\ decays in front of the target
show a similar event topology. 
However, due to the small \ximin\ content of the beam
(see section \ref{sec_2_3_1}) and the decay of $\Lambda^0$ in flight upstream
of the target, the content of $\Lambda^0$ in the neutron beam amounts to
less than 1.26~\%.
On the basis of quark counting rules, we do not expect that the 
$\Lambda^0(uds) + A \rightarrow \Xi^-(dss) + X $ cross section exceeds the
one of the $\Sigma^-(dds) + A \rightarrow \Xi^-(dss) + X $ reaction.
Based on the cross section measured in the present experiment, we find that
in the $x_F$ region of interest ($x_F \leq 0.6$) corrections due to 
$\Lambda^0$ interactions reach at most 22~\% at large $x_F$. For
small $x_F \approx 0$ this contamination falls below 2~\%. 
Since the $\Lambda^0$ cross section is not known, we attribute 
these potential corrections to the systematic uncertainties.

We also estimated the probability for events with two beam particles, where
one particle was not reconstructed and thus giving the same topology as 
a neutron interaction. We therefore scanned our data for events with a 
beam \sigmamin\ reconstructed in the hadron calorimeter and 
an interaction vertex without a beam track pointing to it.
The contamination of the neutron data sample from such type of events was 
found to be negligible.

\subsection{Detection Efficiencies}
\label{sec_2_4}

The large aperture of the spectrometer provided a relatively flat
geometrical acceptance within the region of 
\xf\ $\geq$ {0.05} and \pt\ $\leq$~2.5 \gevc1 .
However, detailed Monte Carlo calculations are required 
in order to determine the reconstruction efficiency for the 
decaying hyperons. 
The primary \ximin\ momentum distribution was generated following a
kinematical distribution of the form
\begin{equation} 
\frac{d^2\sigma}{dp^2_t dx_F} = C (1 - x_F)^n \cdot exp (-bp_t^2).
\end{equation}
For each generated \ximin, the remaining momentum 
was assigned to a virtual \pimin\ which was fed as a beam particle into 
the FRITIOF simulation package \cite{fritiof}
to generate the correlated hadronic background. A complete detector
simulation was performed in the  frame of GEANT \cite{geant}.
These simulations took all available information on the detection 
efficiencies of the individual components of the tracking system
into account.
The simulated events were subsequently passed through the event 
reconstruction chain described above.
To check if the Monte Carlo corrections are self consistent, we
produced two independent sets of events with different $x_F$ and
$p^2_t$ parametrization. We were able to restore the generated $x_F$ and
$p^2_t$ distributions of the first set applying the corrections
obtained in the second set.

The overall detection efficiency
for \ximin\ decays reaches a maximum of about 6\% at  $x_F =
0.25 $. Towards low $x_F \approx 0$ the efficiency 
decreases to about 2.5\% mainly due to acceptance losses for 
large-angle tracks and to the smaller reconstruction efficiency for 
low momentum tracks. 
At high $x_F \approx 1$ the efficiency drops to 1.1\%  
because of the finite fiducial volume for \ximin\/\lambdazero decays. 
According to the simulation, no significant difference in the 
detection efficiency is expected for 
\ximin\ produced in the carbon and copper targets. 

The uncertainty in the efficiency determination is the major source 
of systematic error in the cross section measurements. 
To estimate  this uncertainty we varied the event selection criteria 
applied to the data and the simulated events. The resultant 
changes in the  cross sections are about 20\%.

Finally, uncertainties of the incident beam flux were estimated by subdividing
the data into subsamples and normalizing them individually. We found maximum
variations of 15\%. 

Adding all systematic errors quadratically we derive at
a total systematic uncertainty of 25\% for the \pimin and
\sigmamin\ measurements. For the neutron beam, the systematic
uncertainty rises with $x_F$ from 26\% at $x_F = 0$ to about
33\% at $x_F = 1$.
 
\section{Results}
\label{sec_3}
\subsection{Differential cross sections}
\label{sec_3_1}

The number of observed $\Xi^-$ decays was translated into a cross
section using the formula:
\begin{equation} 
      {  \sigma(x_F,p_t^2)  } = 
            {{ 1 } \over { BR(\Lambda^0 \rightarrow p \pi^-)}} \cdot
            {{ N_{\Xi^-} } \over \varepsilon(x_F,p^2_t) \ N_{b}\ \rho\ l\ N_A / M }.
\end{equation}
Here $N_{\Xi^-}$  is the number of observed \ximin\ , $\varepsilon$ denotes  
the efficiency (including the trigger efficiency) and
acceptance, $ N_{b}$ is the
number of incoming beam particles corrected for the corresponding 
beam contaminations and losses due to the dead time of the
trigger and the data acquisition system.
The target has an atomic mass $M$, a density 
$\rho$, and a geometric length $l$.
$N_A$  stands for the Avogadro number and $ BR(\Lambda^0 \rightarrow p \pi^-) $
accounts for the branching ratio of the observed $\Lambda^0$ decay mode.

\begin{figure}[H]
\begin{center}
\mbox{\epsfxsize=13cm\epsffile{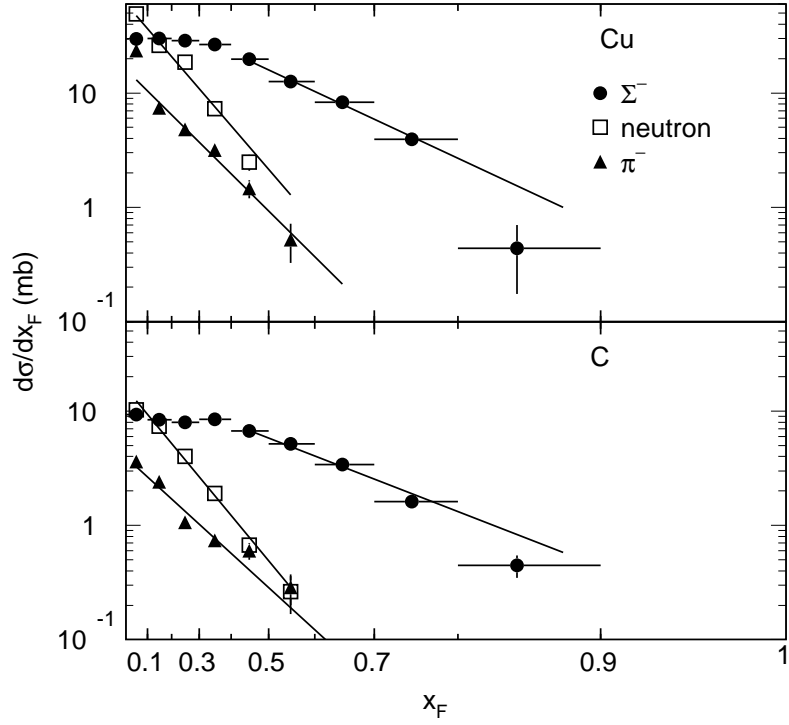}}
\caption{The differential cross section $d\sigma / dx_F$ as a function 
         of $x_F$ integrated
         over $p^2_t$ for neutron, \pimin\ and \sigmamin\ interactions
         with copper (top part) and carbon (bottom part). 
         The lines represent fits proportional to
         $(1-x_F)^n $ with the parameters n as given in table  
         \protect\ref{tab:ksy_crosst}.
         }
\label{fig:ksy_xflog}
\end{center}
\end{figure}

\begin{figure}[H]
\begin{center}
\mbox{\epsfxsize=13cm\epsffile{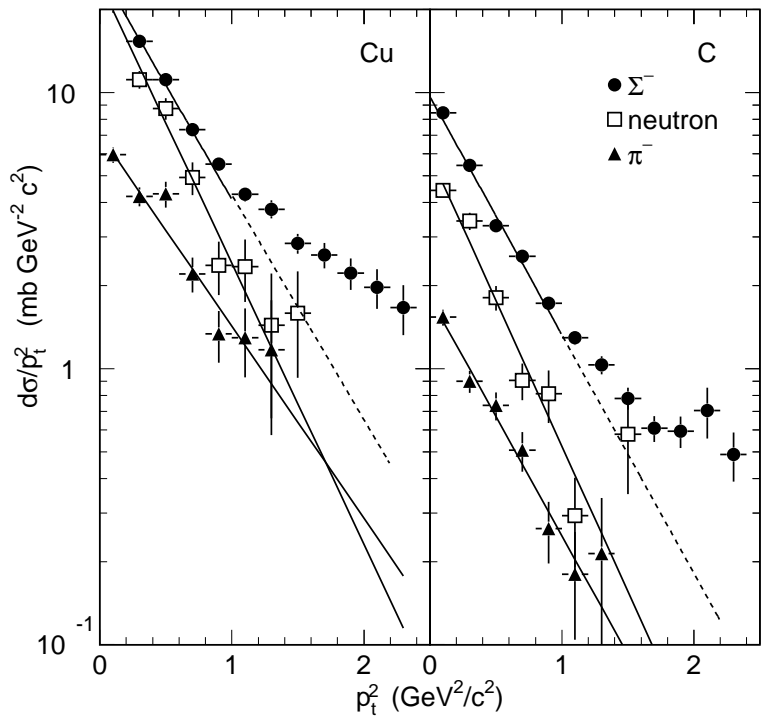}}
\caption{The differential cross section as a function of $p^2_t$ integrated
         over $x_F$ for neutron, \pimin\ and \sigmamin\ interaction
         with copper (left part) and carbon (right part). 
        The lines represent fits proportional to  
         $ exp(-b p_t^2) $; the parameters b are listed in 
         table \protect\ref{tab:ksy_crosst}.
        }
\label{fig:ksy_pt}
\end{center}
\end{figure}

Figures \ref{fig:ksy_xflog} and \ref{fig:ksy_pt} 
display the differential cross sections as a function
of the Feynman variable, $x_F$, and the squared
transverse momentum, $p_t^2$, for the three different projectiles 
and the copper and the carbon targets, respectively. 
Only statistical errors are shown.
The corresponding numbers are listed in 
tables~\ref{tab:ksy_diffxf}~-~\ref{tab:ksy_diffpt}. 

As a point of reference, we parametrized the observed
invariant cross sections by a function of the form
\begin{equation} 
\frac{d^2\sigma}{dp^2_t dx_F} = C (1 - x_F)^n \cdot exp (-bp_t^2)
\end{equation}
which is based on quark counting rules \cite{BLA74} and
phase space arguments.
The three parameters C, b, and n were assumed to be independent of
$p_t$ and $x_F$. 
In case of \sigmamin\ interactions this functional form 
parametrizes the data only in the regions of $x_F> 0.4$ and 
$p^2_t < 1 \Gev2c2 $.
Fits within this limited range are shown by the solid lines 
in Figs. \ref{fig:ksy_xflog} and \ref{fig:ksy_pt}. 
The values of the fit parameters are listed in Table~\ref{tab:ksy_crosst}.
To obtain the total production cross section we integrated the 
differential cross sections in the 
region $0 < x_F < 1$. For neutron and \pimin\ interactions the 
fits shown in Figs. \ref{fig:ksy_xflog} and \ref{fig:ksy_pt} were used 
to extrapolate from the measured range up to $x_F$ =1.
This cross section is also listed in Table~\ref{tab:ksy_crosst}.

The distributions shown in Figure \ref{fig:ksy_xflog} signal a
strong leading effect for the \sigmamin\ projectile at large $x_F$. 
In this projectile fragmentation region the quark overlap between projectile 
and produced particle is reflected in the production yield.
In turn, in the central region the different beam particles yield comparable
\ximin\ cross sections thus indicating that  
the initial strangeness content in the projectile 
is not relevant and all strange quarks are produced in the fragmentation process.

The transverse momentum distributions are -- for $p_t \leq$ 1 \gevc1 -- 
similar for the different projectiles. The most striking feature 
in Figure \ref{fig:ksy_pt} is a strong enhancement at large transverse momenta 
for the \sigmamin\ data as compared to the \Pn\ and \pimin data. 
The slope of this high momentum tail is about half of 
the one describing the low $p_t$ regime, a phenomenon which is consistent with 
early observations at the CERN-ISR \cite{BAN73,CRO75}. It is an
indication that at $p^2_t \approx 1 \Gev2c2 $ higher order double
scattering processes become important. 

For the high statistics \sigmamin\ data we present the invariant double differential 
cross sections in Tables~\ref{tab:ksy_inv1}~-~\ref{tab:ksy_inv2} corrected for the 
different components of our beam. The error bars contain the statistical error 
contribution from data and Monte Carlo.

\subsection{{\sl A} dependence}
\label{sec_3_2}

\begin{figure}[H]
\begin{center}
\mbox{\epsfxsize=13cm\epsffile{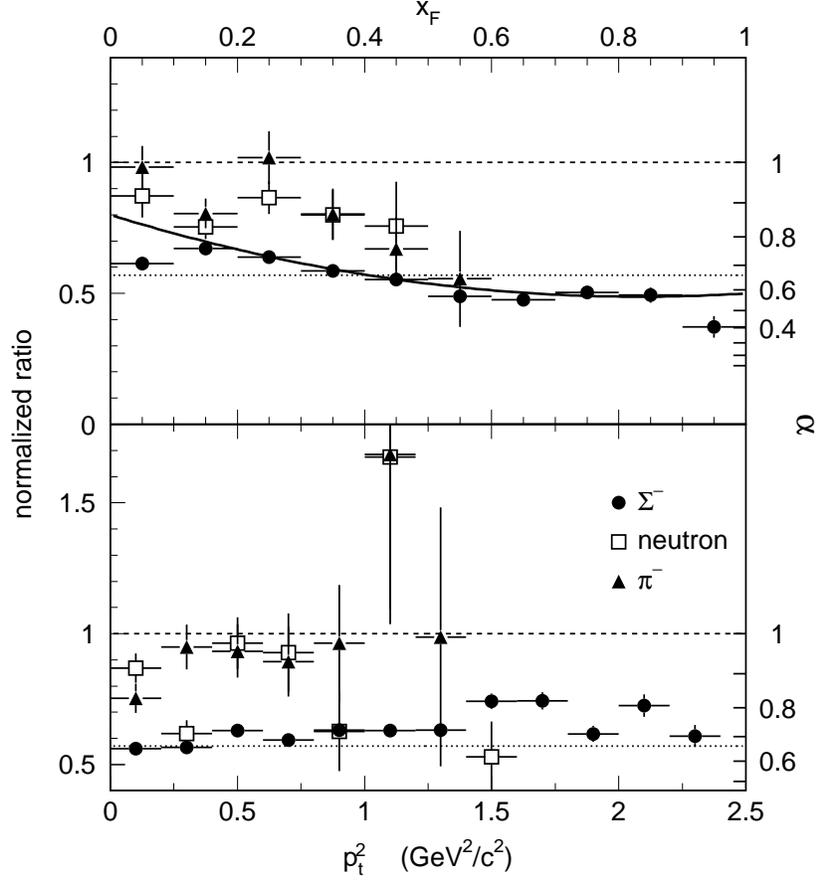}}
\caption
{Normalized ratio $R$ (see Eq. \protect\ref{eq:ratio}) 
of the \ximin\ cross section measured in reactions
with carbon and copper nuclei as a function 
of $x_F$ (top part) and $p^2_t$ (bottom part) for 
neutron, \pimin\, and \sigmamin\ induced reactions.
The right scale indicates the exponent $\alpha$ in case of an 
$A^{\alpha}$ dependence.
The solid line in the upper part marks a polynomial fit to
a compilation of target attenuation factors given in ref. \protect\cite{GEI91}.
}
\label{fig:ksy_alfaxf}
\end{center}
\end{figure}

Acceptance and efficiency corrections for the detected \ximin\ yield do 
not depend on the target.
Therefore the ratio of the cross sections for the copper 
and carbon targets allows the determination of the  
dependence of \ximin\ production on the nuclear mass {\sl A}  with negligible 
systematic uncertainties.
Figure~\ref{fig:ksy_alfaxf} shows the variation of the 
normalized cross section ratio
\begin{equation} 
R=\frac{\sigma_{Cu}}{\sigma_C} \cdot \frac{A_C}{A_{Cu}}
\label{eq:ratio}
\end{equation} 
as a function of $x_F $ (top) and $p^2_t$ (bottom).
Here, the indices C and Cu refer to the carbon and copper target, 
respectively. Using the conventional parametriszation for the {\sl A} dependence
\begin{equation} 
      \sigma = {\sigma}_0 \cdot A^{\alpha} 
\end{equation}
the ratio {\sl R} can be directly translated into the exponent $\alpha$ 
(righthand scale in Fig.~\ref{fig:ksy_alfaxf}).
The average values of this attenuation factor $\alpha$
together with the extrapolated inclusive cross section per nucleon are
listed in Table~\ref{tab:ksy_crossn}.

A compilation of particle production yields in proton - nucleus
collisions given in ref. \cite{GEI91}
suggested for $x_F >$0 a dependence of $\alpha$ on $x_F$ of the form
$\alpha (x_F) = 0.8-0.75x_F+0.45x_F^2$.
While the attenuation for \ximin\ production 
in \sigmamin\ induced interactions is well described by this
parametrization 
(solid line in the upper part of Figure~\ref{fig:ksy_alfaxf}),
the neutron and pion data show significantly larger
values for $\alpha$ at low $x_F$. This deviation from the general behaviour
is in line with the target mass dependence observed 
earlier for $\Xi^0$ production in proton induced reactions at
400 \gevc1 \cite{BER86,GEI91,BAR83}.
It seems that the production of two additional strange 
quarks at small $x_F$ is generally related to a weaker attenuation.   
Within the constituent quark model such a behaviour can be explained
by the assumption that a recombination of leading quarks with a heavy
strange quark is suppressed compared to the recombination with a light {\it up}
 or {\it down} quark \cite{TAK83}.

\section{Comparison with other data}
\label{sec_4}

\begin{figure}[H]
\begin{center}
\mbox{\epsfxsize=13cm\epsffile{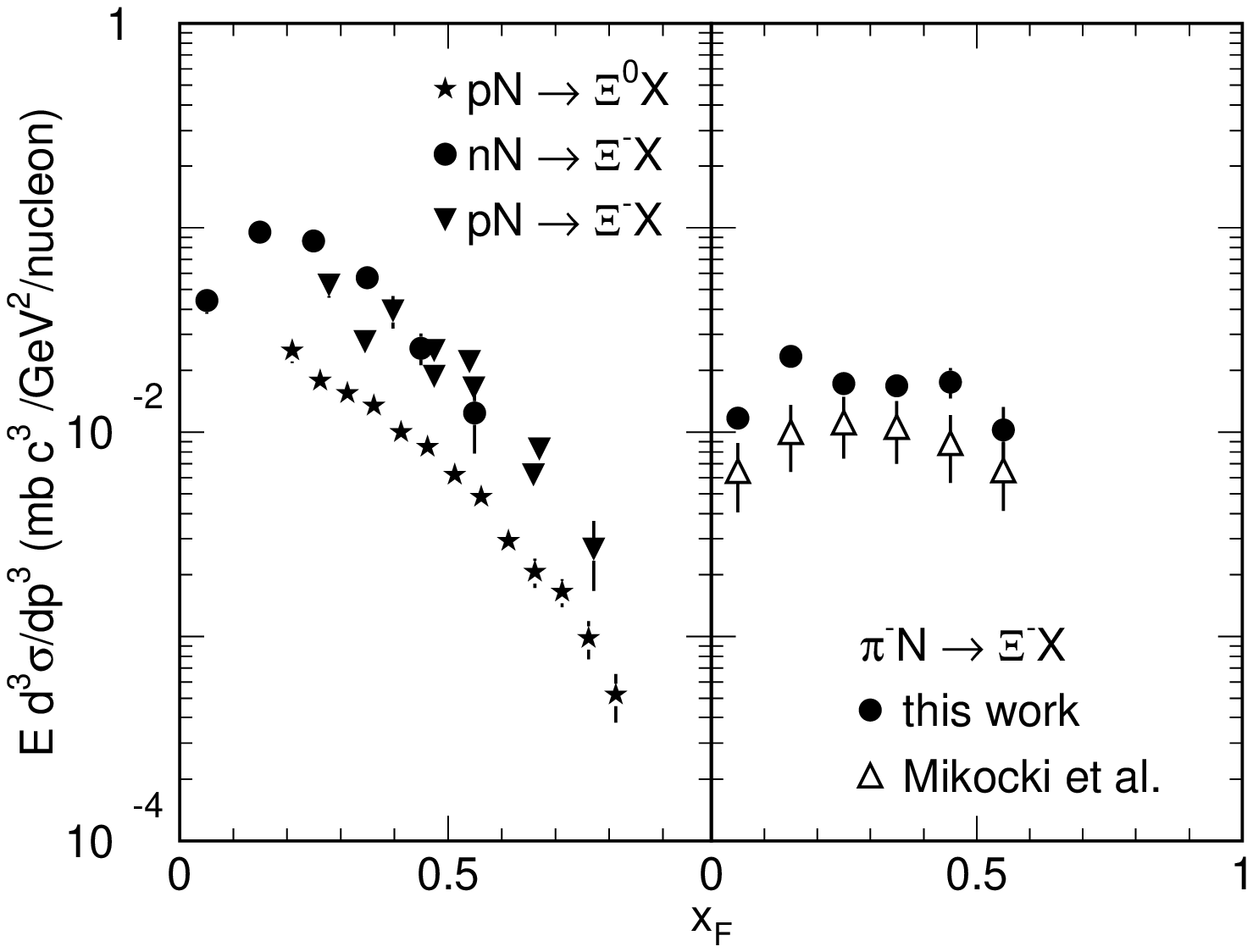}}
\vspace{0cm}
\caption{
Left part: Invariant cross sections for inclusive \ximin\ and 
$\Xi^0$ production at $p_t^2 < 0.2 \ \Gev2c2 $
by protons and neutrons: 
$\Pn A \rightarrow \Xi^-X$ at 345 \gevc1 this measurement (filled circles);
$\Pp A \rightarrow \Xi^-X$ at 200 \gevc1 and 400 \gevc1 
\protect\cite{BOU79,CAR85} (triangles);
$\Pp A \rightarrow \Xi^0X$ at 400 \gevc1\ \protect\cite{BER86} (stars).
The right part compares the invariant inclusive  
$\Xi^-$ production cross sections at $p_t^2 <
0.2 \ \Gev2c2 $ by pions
observed in the present experiment at 345 \gevc1 (filled circles),  to 
previous data taken at 200 \gevc1 (open triangles) \protect\cite{MIK86}.}
\label{fig:compa1}
\end{center}
\end{figure}

In the left part of figure~\ref{fig:compa1} we compare the invariant cross 
section per nucleon measured in neutron induced reactions with data from proton
induced reactions at similar energies. 
As naively expected from the larger d-quark content in the incoming neutron, 
the neutron data lie slightly above the cross section for $pA$ collisions.

The invariant cross section for $\Xi^0$ production in \Pp-nucleon collisions is
given by the stars in Fig.~\ref{fig:compa1} (left). Here, the \Pp-Be data from 
ref. \cite{BER86} were extrapolated to
a nucleon target via a power-law {\sl A} dependence of the form $A^{0.8}$.  
Assuming isospin symmetry, identical cross sections are expected for the 
$ \Pn (udd)+A \rightarrow$ \ximin\ (dss) and the $\Pp (uud)+A \rightarrow$
$\Xi^0 (uss)$ 
 reactions. 
However, the present  $nA \rightarrow$ \ximin\ X cross section is about a factor
of 5 above the $\Pp A \rightarrow$ $\Xi^0$ X data from ref. \cite{BER86}.
At present, no explanation for this unexpected inconsistency exists.

For pion induced \ximin\ production we observe a cross section  
which is about a factor 2 higher than those measured previously in
$\pi+A$ collisions at 200 \gevc1 \cite{MIK86}
(right part of figure~\ref{fig:compa1}). 
In contrast to this old measurement the present values are only about 
50\% smaller than the corresponding cross section for proton induced 
reactions. This difference is close to the ratio of the 
total inelastic cross sections for \pimin and protons. 
Finally, we note that our value for the exponent n describing the 
$x_F$ dependence is only about half of that observed earlier for
$\Xi^{\pm}$ production (n=6.7$\pm$0.3) in ref. \cite{FEN84}.

To check our possible systematic errors we performed a cross section 
measurement for \lambdazero\ , \antilambda\ and \kzero\ in pion and 
neutron interactions. These cross section values
will be the subject of a forthcoming paper. They are in agreement
with the existing world data within 40\% of its values and 
show us that there we don't have additional unknown systematic errors big 
enough to explain the observed difference between the $\Pp A \rightarrow
\Xi^0$ and  $\Pp A \rightarrow \Xi^-$ production cross sections.

\begin{figure}[H]
\begin{center}
\mbox{\epsfxsize=13cm\epsffile{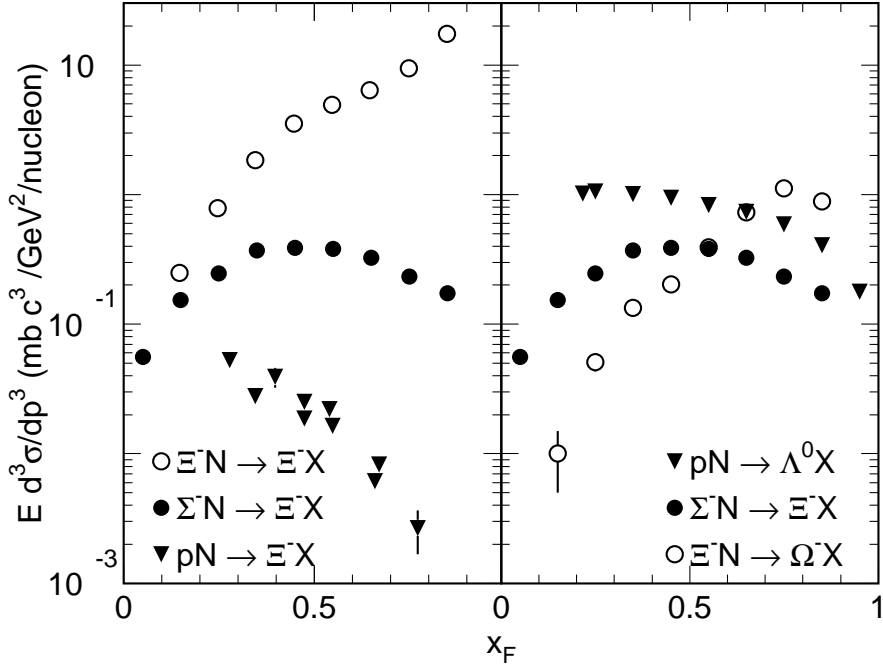}}
\vspace{0cm}
\caption{
Invariant cross sections per nucleon for baryons 
containing strange quarks. The left part displays the \ximin\ cross 
section at $p_t^2 < 0.2 \ \Gev2c2 $ in reactions with 
$\Delta$S=0 (open circles),$\Delta$S=1 (filled circles), 
and $\Delta$S=2 (triangles) between beam and observed particle.
The invariant inclusive particle production cross section at $p_t^2
<0.2 \ \Gev2c2 $ 
in reactions with $\Delta S=1$ between projectiles and observed final particle 
for different projectiles is shown on the right hand side. The data are from
$\Xi^-A\rightarrow\Xi^-X$ and
$\Xi^-A\rightarrow\Omega^-X$ at 116 \gevc1  \protect\cite{BIA87},
$\Sigma^-A\rightarrow\Xi^-X$ at 345 \gevc1 (this measurement),
$\Pp A\rightarrow\Xi^-X$ at 200 \gevc1 and at 400 \gevc1 \protect\cite{BOU79,CAR85},
 and $\Pp A\rightarrow\Lambda^0X$ at 300 \gevc1 \protect\cite{SKU78}.
}
\label{fig:compa}
\end{center}
\end{figure}

In  fig. ~\ref{fig:compa} we compare our data on \ximin\
production by \sigmamin\ with other production cross sections of
hyperons by different projectiles.
Figure ~\ref{fig:compa} (left part) shows the $x_F$ dependence of \ximin\
production by protons, \sigmamin\ and \ximin . Although projectile
and produced \ximin\ differ in strangeness by different amounts the three
cross sections are equal at small $x_F$. At large $x_F$ strong leading effects
are observed for \ximin\ as well as \sigmamin\ projectiles.
At $x_F$ close to unity, each unit of strangeness in the incoming projectile 
causes a cross-section increase by about two orders of magnitude.

A complementary behaviour is observed in the right part of Figure 
~\ref{fig:compa} which compares three reactions, where the incoming  
projectile and the produced
particle differ by one unit of strangeness. Here the cross sections are 
of approximatly equal size at large $x_F$, but differ by about one order
of magnitude at small $x_F$. Together the two figures indicate that in 
the central region the initial strangeness content in the projectile 
is not relevant for the production of \ximin\ and all strange 
quarks contained in the \ximin\ are produced in the fragmentation process.
On the other hand, in the projectile fragmentation region the  cross sections 
depend strongly on the overlap of the quark content between the projectile and 
the produced particle. 
 

\section{Summary and conclusions}
\label{sec_5}


We have presented the first measurement of $\Xi^-$ production by 
$\Sigma^-$, at a beam energy of 345 GeV/c and using carbon and copper targets.
The measurement was supplemented by measurements of $\Xi^-$ production
by neutrons and $\pi^-$, with lower statistics.

At $p_t$ values above 1 GeV/c, the cross-section for $\Xi^-$ production
by $\Sigma^-$ shows a marked increase compared to the 
usually observed Gaussian behaviour of $d\sigma / dp_t^2$,
which indicates that a different production process becomes important.

The A-dependence of the $\Xi^-$ production cross-section
is close to a value of $\alpha=2/3$ for production by $\Sigma^-$,
while the less precise data on production by neutrons and $\pi^-$
favour a higher value $\alpha=0.9$.
All three cross-sections  show a trend to decreasing values of $\alpha$
with increasing $x_f$.

Difficulties encountered in comparing absolute cross-sections from
different experiments  with pion and neutron beams
underline the necessity of  comprehensive high-statistics studies
of particle production within one experiment
for more detailed investigations of hadronic interactions.

Nonetheless, a comparison of our result on the differential cross-section $d\sigma /dx_F$ for
$\Xi^-$ production by
$\Sigma^-$
with existing data on hyperon production by nucleons and $\Xi^-$
shows a strong leading particle effect:
at low $x_F$ the baryon production cross-sections depend strongly on the
strangeness of the produced particle, and are independent of the
quark flavour of the beam particle. At high  $x_F$, on the other hand,
the cross-sections depend strongly on the strange quark overlap 
between beam particle and produced particle, i.e.  the cross-sections at
high  $x_F$ are strongly enhanced by quark transfer from the beam particle
to the outgoing particle.
These results emphasize the power of experiments 
with beams of different strangeness to distinguish between 
different contributions to the complex phenomenon
of inclusive hadronic production.

\section*{ Acknowledgements }

We are indebted to J.Zimmer and the late Z.Kenesei
for their help during all moments of detector construction and set-up.
We are grateful to the staff of CERN's 
EBS group for providing an excellent
hyperon beam channel, to the staff of CERN's Omega group for their 
help in running
the $\Omega $-spectrometer and also to the staff of the SPS for providing
good beam conditions. 
We thank D.M.Jansen, B.Kopeliovich, R.Ransome and O.Piskounova for many 
helpful discussions.

\newpage
\begin{table}[H]
\begin{center}
\begin{tabular}{|c|r|c|r|c|c|c|} \hline
Beam& \multicolumn{2}{c|}{Neutrons} & \multicolumn{2}{c|}{$\pi^{-}$} &
                        \multicolumn{2}{c|}{$\Sigma^{-}$}  \\ \hline
$x_F$&\multicolumn{1}{c|}{Copper}&Carbon&\multicolumn{1}{c|}{Copper}
     &Carbon&Copper&Carbon \\ \hline
0.0 - 0.1 & $ 49.33 \pm  6.50 $ & $ 10.26 \pm  1.40 $ &
            $ 23.59 \pm  2.88 $ & $  3.61 \pm  0.41 $ &
            $ 29.82 \pm  1.11 $ & $  9.30 \pm  0.33 $ \\
0.1 - 0.2 & $ 26.22 \pm  1.81 $ & $  7.38 \pm  0.55 $ &
            $  7.39 \pm  0.57 $ & $  2.41 \pm  0.20 $ &
            $ 30.01 \pm  0.66 $ & $  8.39 \pm  0.20 $ \\
0.2 - 0.3 & $ 18.57 \pm  1.33 $ & $  4.02 \pm  0.31 $ &
            $  4.81 \pm  0.43 $ & $  1.06 \pm  0.11 $ &
            $ 28.80 \pm  0.65 $ & $  7.96 \pm  0.18 $ \\
0.3 - 0.4 & $  7.32 \pm  0.70 $ & $  1.90 \pm  0.20 $ &
            $  3.17 \pm  0.37 $ & $  0.74 \pm  0.09 $ &
            $ 26.46 \pm  0.64 $ & $  8.47 \pm  0.21 $ \\
0.4 - 0.5 & $  2.49 \pm  0.40 $ & $  0.67 \pm  0.12 $ &
            $  1.46 \pm  0.26 $ & $  0.60 \pm  0.10 $ &
            $ 19.74 \pm  0.53 $ & $  6.71 \pm  0.19 $ \\
0.5 - 0.6 & \multicolumn{1}{c|}{$-$} & $  0.26 \pm  0.10 $ &
            $  0.52 \pm  0.19 $ & $  0.29 \pm  0.08 $ &
            $ 12.58 \pm  0.40 $ & $  5.16 \pm  0.17 $ \\
0.6 - 0.7 & \multicolumn{1}{c|}{$-$}& \multicolumn{1}{c|}{$-$}&
            \multicolumn{1}{c|}{$-$}& \multicolumn{1}{c|}{$-$}&
            $  8.32 \pm  0.37 $ & $  3.40 \pm  0.14 $ \\
0.7 - 0.8 & \multicolumn{1}{c|}{$-$}& \multicolumn{1}{c|}{$-$} &
            \multicolumn{1}{c|}{$-$}& \multicolumn{1}{c|}{$-$}&
            $  3.93 \pm  0.29 $ & $  1.61 \pm  0.11 $ \\
0.8 - 0.9 & \multicolumn{1}{c|}{$-$}& \multicolumn{1}{c|}{$-$}&
            \multicolumn{1}{c|}{$-$}& \multicolumn{1}{c|}{$-$}&
            $  0.44 \pm  0.26 $ & $  0.45 \pm  0.10 $ \\
0.9 - 1.0 & \multicolumn{1}{c|}{$-$}& \multicolumn{1}{c|}{$-$}&
            \multicolumn{1}{c|}{$-$}& \multicolumn{1}{c|}{$-$}&
            \multicolumn{1}{c|}{$-$}& \multicolumn{1}{c|}{$-$} \\
\hline
\end{tabular}
\end{center}
\caption{ Differential cross section of \ximin\ production 
          as a function of $x_F$ in mb.}
\label{tab:ksy_diffxf}
\end{table}
\begin{table}[h]
\begin{center}
\begin{tabular}{|c|c|c|c|c|r|c|} \hline
Beam& \multicolumn{2}{c|}{Neutrons} & \multicolumn{2}{c|}{$\pi^{-}$} &
                        \multicolumn{2}{c|}{$\Sigma^{-}$}  \\ \hline
$p_t^2$&Copper&Carbon&Copper&Carbon&\multicolumn{1}{c|}{Copper}
       &Carbon \\ \hline

0.0 - 0.2 & $ 20.53 \pm  1.07 $ & $  4.43 \pm  0.25 $ &
            $  5.96 \pm  0.39 $ & $  1.54 \pm  0.10 $ &
            $ 24.84 \pm  0.40 $ & $  8.45 \pm  0.13 $ \\
0.2 - 0.4 & $ 11.15 \pm  0.82 $ & $  3.42 \pm  0.25 $ &
            $  4.21 \pm  0.33 $ & $  0.90 \pm  0.08 $ &
            $ 15.37 \pm  0.34 $ & $  5.44 \pm  0.12 $ \\
0.4 - 0.6 & $  8.76 \pm  0.78 $ & $  1.81 \pm  0.18 $ &
            $  4.29 \pm  0.46 $ & $  0.74 \pm  0.09 $ &
            $ 11.14 \pm  0.32 $ & $  3.30 \pm  0.10 $ \\
0.6 - 0.8 & $  4.92 \pm  0.67 $ & $  0.91 \pm  0.14 $ &
            $  2.21 \pm  0.32 $ & $  0.51 \pm  0.08 $ &
            $  7.35 \pm  0.27 $ & $  2.56 \pm  0.10 $ \\
0.8 - 1.0 & $  2.37 \pm  0.52 $ & $  0.81 \pm  0.18 $ &
            $  1.34 \pm  0.28 $ & $  0.26 \pm  0.07 $ &
            $  5.52 \pm  0.25 $ & $  1.73 \pm  0.08 $ \\
1.0 - 1.2 & $  2.34 \pm  0.59 $ & $  0.29 \pm  0.11 $ &
            $  1.29 \pm  0.36 $ & $  0.18 \pm  0.08 $ &
            $  4.29 \pm  0.25 $ & $  1.29 \pm  0.08 $ \\
1.2 - 1.4 & $  1.44 \pm  0.78 $ & $  0.00 \pm  0.00 $ &
            $  1.17 \pm  0.60 $ & $  0.21 \pm  0.13 $ &
            $  3.78 \pm  0.28 $ & $  1.03 \pm  0.08 $ \\
1.4 - 1.6 & $  -              $ & $  -              $ &
            $  -              $ & $  -              $ &
            $  2.84 \pm  0.24 $ & $  0.78 \pm  0.07 $ \\
1.6 - 1.8 & $  -              $ & $  -              $ &
            $  -              $ & $  -              $ &
            $  2.58 \pm  0.27 $ & $  0.61 \pm  0.07 $ \\
1.8 - 2.0 & $  -              $ & $  -              $ &
            $  -              $ & $  -              $ &
            $  2.22 \pm  0.29 $ & $  0.59 \pm  0.08 $ \\
2.0 - 2.2 & $  -              $ & $  -              $ &
            $  -              $ & $  -              $ &
            $  1.97 \pm  0.32 $ & $  0.71 \pm  0.15 $ \\
2.2 - 2.4 & $  -              $ & $  -              $ &
            $  -              $ & $  -              $ &
            $  1.67 \pm  0.34 $ & $  0.49 \pm  0.10 $ \\
\hline
\end{tabular}
\end{center}
\caption{ Differential cross section of \ximin\ production 
          as a function of $p_t^2$ in  mb/$(\gevc1)^{-2}$.}
\label{tab:ksy_diffpt}
\end{table}

\newpage

\begin{table}[h]
\begin{center}
\begin{tabular}{|l|l|r|c|l|c|c|c|} \hline
Beam     &Target & \# events & average total &cross section per& n& $b$\\ 
particle &       &           & efficiency    &nucleus~~$(mb)$&~&(\gevc1)$^{-2}$\\ 
 \hline
\sigmamin\ & Copper & $90160 \pm 336 $ & 0.029 &$18.31 \pm 0.09$ &1.97$\pm$0.04&1.90$\pm$0.04\\
           & Carbon & $87528 \pm 331 $ & 0.029 &$ 5.15 \pm 0.03$ &2.08$\pm$0.04&2.00$\pm$0.04\\
 \hline
 Neutrons  & Copper & $ 1847 \pm 47  $ & 0.034 &$10.6 \pm 0.1$ &4.8$\pm$0.3&2.3$\pm$0.2\\
           & Carbon & $ 1627 \pm 44  $ & 0.034 &$ 2.5 \pm 0.1$ &5.0$\pm$0.3&2.4$\pm$0.2\\
 \hline
  \pimin\  & Copper & $ 1683 \pm 48  $ & 0.043 &$ 4.1 \pm 0.1$ &4.1$\pm$0.3&1.6$\pm$0.1\\
           & Carbon & $ 1448 \pm 44  $ & 0.043 &$ 0.9 \pm 0.1$ &3.8$\pm$0.3&1.9$\pm$0.1\\
\hline
\end{tabular}
\end{center}
\caption{ Number of reconstructed events, average efficiency, total 
inclusive \ximin\ production cross sections and parameters 
of fits using a function of the form $ {d^2 \sigma}/{dp^2_t dx_F} = C(1-x_F)^n exp(-bp^2_t) $
for the different beam particles and targets. Only statistical errors are given.
}
\label{tab:ksy_crosst}
\end{table}

\begin{table}[htb]
\begin{center}
\begin{tabular}{|c|c|c|c|c|c|} \hline
  & \multicolumn{4}{c|}{$E \cdot{d^3 \sigma} / {dp^3} \ \ [mb \cdot c^3 
                        \cdot \mathrm{GeV}^{-2} ] $} & \\
$x_F$&$0.0<p_t^2<0.4$&$0.4<p_t^2<0.8$&$0.8<p_t^2<1.2$&$1.2<p_t^2<1.6$ &
      $1.6<p_t^2<2.0$  \\ \hline
0.0 - 0.2 & $  30.2 \pm   0.6 $ & $  15.2 \pm   0.7 $ &
            $  10.8 \pm   0.8 $ & $  10.2 \pm   1.5 $ &
            $   8.9 \pm   2.3 $ \\
0.2 - 0.4 & $  88.3 \pm   1.7 $ & $  42.7 \pm   1.7 $ &
            $  23.6 \pm   1.6 $ & $  15.7 \pm   1.7 $ &
            $  19.4 \pm   4.6 $ \\
0.4 - 0.6 & $  95.8 \pm   2.3 $ & $  46.6 \pm   2.0 $ &
            $  23.2 \pm   1.6 $ & $  17.7 \pm   2.0 $ &
            $  11.2 \pm   2.2 $ \\
0.6 - 0.8 & $  67.5 \pm   2.3 $ & $  34.6 \pm   1.9 $ &
            $  21.7 \pm   2.0 $ & $  17.0 \pm   2.6 $ &
            $   9.0 \pm   2.7 $ \\
0.8 - 1.0 & $  38.8 \pm   2.6 $ & $  12.9 \pm   1.4 $ &
            $   9.7 \pm   2.5 $ & $         -       $ &
            $        -        $ \\
 \hline
\end{tabular}
\end{center}
\caption{ Invariant cross section of \ximin\ production by \sigmamin\
on copper as a function of \xf\ and $p_t^2$.}
\label{tab:ksy_inv1}
\end{table}

\begin{table}[htb]
\begin{center}
\begin{tabular}{|c|c|c|c|c|c|} \hline
  & \multicolumn{4}{c|}{$E \cdot{d^3 \sigma} / {dp^3} \ \ [mb \cdot c^3 
                        \cdot \mathrm{GeV}^{-2} ] $} & \\
$x_F$&$0.0<p_t^2<0.4$&$0.4<p_t^2<0.8$&$0.8<p_t^2<1.2$&$1.2<p_t^2<1.6$ &
     $1.6<p_t^2<2.0$  \\ \hline
0.0 - 0.2 & $   8.9 \pm   0.2 $ & $   4.1 \pm   0.2 $ &
            $   2.8 \pm   0.2 $ & $   2.2 \pm   0.3 $ &
            $   2.3 \pm   0.8 $ \\
0.2 - 0.4 & $  27.5 \pm   0.6 $ & $  12.6 \pm   0.5 $ &
            $   6.2 \pm   0.4 $ & $   3.8 \pm   0.4 $ &
            $   3.9 \pm   0.9 $ \\
0.4 - 0.6 & $  35.8 \pm   0.9 $ & $  15.1 \pm   0.7 $ &
            $   8.8 \pm   0.7 $ & $   6.0 \pm   0.8 $ &
            $   3.3 \pm   0.6 $ \\
0.6 - 0.8 & $  30.2 \pm   1.0 $ & $  10.7 \pm   0.6 $ &
            $   6.9 \pm   0.6 $ & $   5.9 \pm   1.0 $ &
            $   2.7 \pm   1.0 $ \\
0.8 - 1.0 & $  15.9 \pm   1.0 $ & $   8.4 \pm   0.9 $ &
            $   3.2 \pm   0.7 $ & $        -        $ &
            $          -      $ \\
 \hline
\end{tabular}
\end{center}
\caption{ Invariant cross section of \ximin\ production by \sigmamin\
on carbon as a function of \xf\ and $p_t^2$.}
\label{tab:ksy_inv2}
\end{table}

\newpage

\begin{table}[H]
\begin{center}
\begin{tabular}{|l|c|c|} \hline
 Beam           & $\alpha$ & cross section per \\
 particle       &          & nucleon, $mb (x_F>0) $  \\
 \hline
1. \sigmamin\ &$0.681 \pm 0.001$&$ 1.06 \pm 0.01$ \\
2. Neutrons   &$0.880 \pm 0.04$ &$ 0.24 \pm 0.10$  \\
3. \pimin\    &$0.920 \pm 0.03$ &$ 0.11 \pm 0.09$  \\
\hline
\end{tabular}
\end{center}
\caption{ {\sl A} dependence of the inclusive cross section of the \ximin\ production.}
\label{tab:ksy_crossn}
\end{table}


\newpage

\newpage

\noindent
\newpage
\noindent
\newpage
\noindent

\pagebreak
\noindent
\vspace{-1cm}
\noindent
\newpage

\noindent

\newpage
\vspace{-6cm}

\vspace{-1.5cm}
\noindent

\normalsize

\end {document}